\documentclass[journal]{IEEEtran}
\usepackage[cmex10]{amsmath}
\usepackage{exscale,times}
\usepackage[T1]{fontenc}
\usepackage{amssymb}
\usepackage{color}
\usepackage{cite}
\usepackage{graphicx}
\usepackage{epsfig}
\usepackage{multirow}
\usepackage{float}
\usepackage{caption}
\usepackage{subcaption}
\usepackage{url}
\usepackage{dblfloatfix}

\begin{document}

\title{Design, Implementation and Evaluation of \\ a Variable Stiffness Transradial Hand Prosthesis}

\author{Elif~Hocaoglu,~\IEEEmembership{Member,~IEEE}
        and~Volkan~Patoglu,~\IEEEmembership{Member,~IEEE}
\thanks{E. Hocaoglu and V. Patoglu are with the Faculty of Engineering and Natural Sciences of Sabanc\i\ University, \.Istanbul, Turkey.
       {\tt\footnotesize \{elifhocaoglu,vpatoglu\}@sabanciuniv.edu}}%
\thanks{Manuscript received October 26, 2019; revised November XX, 2019.}}

\markboth{IEEE/ASME TRANSACTIONS ON MECHATRONICS, VOL. XX, NO. X, OCTOBER 2019}%
{Shell \MakeLowercase{\textit{et al.}}: Bare Demo of IEEEtran.cls for Journals}

\maketitle

\begin{abstract}
We present the design, implementation, and experimental evaluation of a low-cost, customizable, easy-to-use transradial hand prosthesis capable of adapting its compliance.  Variable stiffness actuation (VSA) of the prosthesis is based on antagonistically arranged tendons coupled to nonlinear springs driven through a Bowden cable based power transmission.  Bowden cable based antagonistic VSA can, not only regulate the stiffness and the position of the prosthetic hand, but also enables a light-weight and low-cost design, by opportunistic  placement of motors, batteries and controllers on any convenient location on the human body, while nonlinear springs are conveniently integrated inside the forearm. The transradial hand prosthesis also features tendon driven underactuated compliant fingers that allow natural adaption of the hand shape to wrap around a wide variety of object geometries, while the modulation of the stiffness of their drive tendons enables the prosthesis to perform various tasks with high dexterity. The compliant fingers of the prosthesis add inherent robustness and flexibility, even under impacts. The control of the variable stiffness transradial hand prosthesis is achieved by an sEMG based natural human-machine interface.
\end{abstract}


%
\IEEEpeerreviewmaketitle

\section{Introduction}
\label{Sec:Introduction}

\IEEEPARstart{V}{ersatile} grasping and manipulation in unstructured environments are challenging tasks actively investigated in robotics. Multi-fingered robot hands  have been developed both in academia~\cite{AnthropomorphicRobHand,ContinualGrasping,ExtrinsicActuation} 
and for commercial use~\cite{ilimb,Ottobockgripper,bostonarm} 
to achieve various tasks. Anthropomorphism (ability to  emulate human-like hand shape, size, and consistency) and dexterity (successful manipulation capability even under unstructured conditions) are commonly identified as  the key features to reach a satisfactory level of performance.

Anthropomorphism is an important criteria in the design of robotic end-effectors, especially for the purpose of hand prostheses~\cite{Bicchi2000,Robotics2008}, since the tools around the environment, e.g., consoles, handles, keys, are designed for the human hands. In addition, anthropomorphic designs are aesthetically and physiologically more fulfilling for amputees, as they provide more natural appearances. However, anthropomorphism alone is not sufficient;  other important criteria, such as simple but robust design, ease of use, and adequate level of dexterity  are also crucial factors in the design of prosthetic hands.

Dexterity is a quite evident goal for the robotic and prosthetic hands in order for them to be endowed with human-like capabilities, such as grasping objects and performing fine finger movements for precise manipulations. In order for a prosthetic hand to qualify as a dexterous design, it has to be capable of performing most of the human hand taxonomy required during the activities of daily living (ADL)~\cite{dollar2016}. In the literature~\cite{Bicchi2011,Bicchi2012}, it has been emphasized that the majority of human grasps are power grasps, which is preferred for more than $50\%$ of the time when the hand is used. Pinch grasp is ranked as second with a $20\%$ preference rate~\cite{dollar2016}. Hence, since power grasps and pinch grasps are the most common hand functions, providing hand prostheses with these dominant grasp types may be sufficient to execute most ADL.

Successful manipulation  necessitates another significant and commonly neglected characteristics of human hand, namely the impedance modulation. Incorporating impedance modulation property in the design of a hand prosthesis makes it adaptable to interacted objects/tasks. Successful execution of many ADL, where human physically interacts with the environment, arises from proper modulation of the impedance level of hand based on the varying requirements of the task. For instance, some activities, such as writing and painting, necessitate highly accurate position control for which the stiffness of the fingers is increased considerably, while manipulation of soft/fragile objects, such as holding an egg or picking up an apricot, requires low stiffness of the fingers.

The impedance modulation property of human hands has inspired design and control strategies for robotic and prosthetic hands, whose goal is to improve quality of interaction with dynamic environments, especially under unpredictable conditions. Specifically, hand prostheses become safer and more functional if the appropriate impedance level based on the physical conditions of the interacted environment can be ensured~\cite{hoganpart0,hoganpart1,hoganpart2}. Recent studies~\cite{Okamura2011,okamura2012,okamura2013} provide strong evidence that hand prosthesis with stiffness modulation can improve the performance of an amputee, when the impedance of the prosthesis is matched to the requirements of the task.

Impedance modulation approaches in robotics can be grouped into two major categories. In the first category, the compliance of the device is modulated through software, using strategies such as impedance/admittance control. In this approach, the impedance modulation is limited by the controllable bandwidth of the actuators, and for this reason, a prosthesis whose impedance is modulated with such a control strategy behaves like a rigid body for high frequency excitations, such as impacts that exceed its control bandwidth~\cite{hoganpart1,hoganpart2}. Moreover, this approach requires continuous use of actuators and suffers from low energy efficiency.

In the second category, the impedance modulation is embedded into the mechanical design of a robotic system. In this approach, impedance of the robotic manipulator is adjusted through special mechanisms consisting of passive elastic elements, such as springs. In hardware based impedance modulation, e.g., variable stiffness actuation~(VSA), the impedance change is physical and is valid for the whole frequency spectrum, including frequencies well above the controllable bandwidth of the actuators. Furthermore, this approach consumes energy only when the impedance is being modulated; hence, is energy efficient.

Achieving human level dexterity  with fully actuated, high degrees of freedom prosthetic hands requires use of complex control algorithms~\cite{Touvet2012}. In the literature, there exists myoelectrically controlled hand prostheses that are capable of performing a large variety of manipulation tasks~\cite{Kuiken,Castellini,Chu}; however, the control of each finger joint in these systems is realized by means of sophisticated learning algorithms, whose complexity exposes amputees to long training periods.  Burden of long training periods and complexity involved in controlling prosthetic devices are known to contribute to high abandonment rate for these devices, reaching up to 40\%~\cite{14}. A large percentage of amputees  reject active prostheses, since they are dissatisfied with the current level of the functionality provided by these devices, given their complexity. These amputees prefer easy to use  passive prostheses, even at a cost of reduced functionality~\cite{15}. To address the  challenges of low adaptation and high abandonment rate of active prostheses, several research groups have focused on the simplification of mechanical design and ensuring ease of control, without loosing the main functionality of prosthetic hands.

In the literature, several robotic hands employed for tasks requiring human machine interaction have been designed using VSA~\cite{DLR1,DLR2,shadow}, while, to the best of authors' knowledge, no such application has been reported in the field of anthropomorphic hand prostheses. In particular, each active degree of freedom of DLR Hand Arm System is controlled by two motors attached to antagonistically arranged nonlinear spring elements~\cite{DLR1,DLR2}. Similarly, the impedance modulation of the anthropomorphic Shadow Hand is achieved by antagonistically arranged pneumatic artificial muscles~\cite{shadow}. Both DLR Hand Arm System and Shadow Hand feature sophisticated mechanical designs with large number of active degrees of freedom; hence, their size, weight and cost make them infeasible for use as a hand prosthesis. Furthermore, grasp planning and impedance modulation of these devices necessitate complex algorithms, which renders their use quite challenging.

Employment of underactuated mechanisms for hand designs is a promising approach, as underactuated hands have been shown to provide a remarkable adaptation to various object geometries  without the need for sensors or complex control algorithms.  Underactuation is commonly implemented by either linkage or tendon based finger designs that are capable of performing typical human-like finger closing sequences.  Linkage based underactuated fingers~\cite{Prattichizzo,Birglen,Laliberte,Kamikawa} are capable of shape adaptation and can endure larger forces compared to tendon based ones; however, their relatively bulky design make them not well-suited for integration into anthropomorphic prosthetic hands. Most successfully implementations of anthropomorphic underactuated hand designs have been realized by means of tendon driven mechanisms~\cite{tendon_driven1,tendon_driven2,tendon_driven_prosthesis1,tendon_driven_prosthesis2}, since slim and lightweight fingers can be actuated with this method. For instance, in~\cite{dollar1,dollar2,dollar3},  a compliant, underactuated, sensor integrated robotic hand with tendon driven elastic joints is introduced and fabricated via support decomposition manufacturing. Thanks to the underactuated finger mechanism with elastic joints, this low cost hand is capable of self-adaptation to different shaped objects under simple control methods.  Similarly, a tendon driven underactuated robot hand that explores synergies of human hand motions is implemented in~\cite{SoftHand}. However, neither of these underactauted hands feature impedance modulation capabilities.

In this study, we present the design, fabrication, and evaluation of a variable stiffness transradial hand prosthesis to be controlled through a natural human-machine interface. Variable stiffness actuation of the prosthesis is based on antagonistically arranged tendons coupled to nonlinear springs driven through a Bowden cable based power transmission. Bowden cable based antagonistic VSA  regulates both the impedance and the position of the hand. It also enables a light-weight hand design, by  opportunistically placing the motors, batteries and controllers to any convenient location on the human body, while nonlinear springs are conveniently integrated inside the forearm. The proposed prosthesis features tendon driven underactuated compliant fingers that enable natural adaption of the hand shape to wrap around a wide variety of object geometries and modulation of hand's stiffness to perform various tasks with high dexterity. The compliant fingers are built from polyurethane with a low-cost  manufacturing process and add inherent robustness and flexibility, even under unexpected conditions such as impacts.

The control of the variable stiffness transradial hand prosthesis is achieved by a natural human-machine interface that utilizes sEMG signals measured from the surface of the upper arm, chest and shoulder. This natural control interface, called \emph{tele-impedance controller}, is first presented in~\cite{hocaoglue12}, while the detailed implementation of this controller and its performance evaluation are presented in~\cite{Part2}.

The rest of the manuscript is organized as follows: Section II-A introduces the design goals for the variable stiffness transradial hand prosthesis, while Section II-B reviews its tele-impedance control architecture detailed in~\cite{Part2}. Section~III details the mechatronic design of the VSA prosthesis. Section VI presents experimental evaluations that provide evidence of the working principle. Section V evaluates the grasping performance of the hand prosthesis with a wide variety of objects and  provides a discussion of the results. Finally, Section VI concludes the paper and presents the future work.

\section{Variable Stiffness Hand Prosthesis}

This section presents the design objectives for VSA hand prosthesis and overviews its sEMG-based control approach.

\subsection{Design Objectives} \label{Sec:DesignGoals}

Following the terminology in~\cite{merlet06}, one can categorize the performance requirements for hand prostheses into four groups as imperative, optimal, primary, and secondary requirements.

Anthropomorphism is an imperative design requirement for hand prostheses. Aesthetically pleasing natural appearance is not only necessary for adaptation of prosthetic devices by amputees, but also anthropomorphic designs are better suited to interact with common human-oriented tools and environments. In this study, we design an anthropomorphic hand prosthesis, for which the dimensions are customizable.

Dexterity is an optimal performance requirement that needs to be maximized while designing a hand prosthesis. In particular, a hand prosthesis should be capable of grasping objects with different shapes (prismatic, spherical, cylindrical) and with various properties (soft or fragile structures, smooth or ragged surfaces) without damaging them. In this study, we ensure dexterity by designing an underactuated hand that can mimic the opening/closing sequence of human fingers to adapt to wide variety of geometries and by enabling the stiffness of the prosthesis to be actively modulated based on the task.

The primary requirement for a hand prosthesis is ease-of-use. The control of the device should be intuitive, allowing amputees to use the device without being exposed to long training periods. Moreover, the hand prosthesis should be energy efficient and its batteries should be easily swappable for user friendliness. In this study, the use of an underactuated design simplifies control of the device, as only position and impedance of the drive tendon needs to be controlled. The tele-impedance controller reviewed in Section II-B provides a natural sEMG interface for the control of the device, where the impedance modulation takes place automatically, allowing the amputee only to focus on the position control of the hand. Energy efficiency is ensured by hardware based impedance modulation, where no energy is wasted to maintain a desired impedance level. Finally, Bowden cable based actuation enables batteries to be opportunistically placed anywhere on the body, making them easily re-sizeable or swappable.

The secondary requirements for the device are high robustness and low cost. In this study, the compliant fingers and the variable stiffness actuation provide built-in physical compliance that provide inherent robustness to impacts. Besides, all parts of the prosthesis are simple to manufacture and customizable. Furthermore, since Bowden cable based actuation allows for motors, drivers and batteries of the system to be remotely located, the cost of these parts can be kept low, as strict size and weight constraints do not apply to these parts.

\vspace{-1.2\baselineskip}

\subsection{Overview of sEMG-based Control Architecture} \label{Sec:Tele-impedance}

\begin{figure}[t]
\centering
\resizebox{3.3in}{!}{\includegraphics{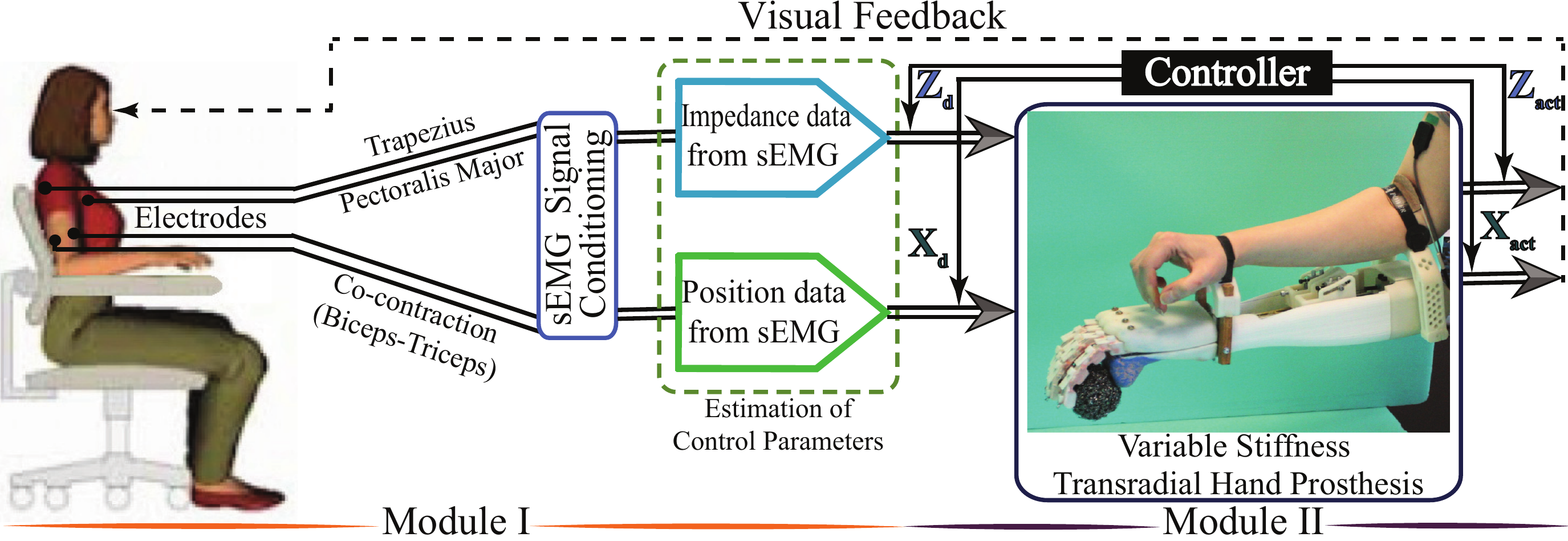}}
\vspace*{-0.5\baselineskip}
\caption{Tele-impedance control architecture of the variable stiffness transradial hand prosthesis.}
\label{teleimpedance}
\vspace*{-.85\baselineskip}
\end{figure}

The tele-impedance control architecture consists of two modules, as depicted in Figure~\ref{teleimpedance}. The first module handles the measurement of sEMG signals, their conditioning, and the estimation of reference values for the hand position and stiffness. The second module implements a closed loop controller that ensures that the position and the stiffness of the VSA prosthetic hand match the reference values. Throughout the control, visual feedback and physical coupling provide information to the amputees to adapt their sEMG signals to match the requirements of the task.

Given that transradial upper extremity amputees lack the muscle groups responsible for the hand and forearm motions, sEMG signals for the position control of the hand prosthesis are measured from the chest and shoulder, while sEMG signals  used for stiffness regulation are measured from the muscle pairs on the upper arm. The estimation of the hand position and stiffness from sEMG signals involves modeling of hand motion/stiffness based on sEMG signals, empirical determination of the parameters of these models for use in real time control, and incorporation of fatigue compensation.

Conditioned sEMG signals are discretized into several levels to map them to the physical stiffness range of the VSA and to the range of motion of the fingers. A calibration procedure is performed for sEMG signals before each use, to customize the maximum voluntary contraction levels for the user, as commonly done for the commercial prosthetic hands.

The sEMG based tele-impedance control interface, together with the VSA, enables an amputee to modulate the stiffness of the prosthetic hand to properly match the requirements of the task, while performing ADL under visual feedback. The regulation of stiffness is managed through the stiffness measurements of
the intact upper arm and this control takes place naturally and automatically as the amputee interacts with the environment. The position of the hand prosthesis is controlled intentionally by the amputee through the  position of the shoulder estimated using SEMG signals. This natural human-robot interface is advantageous, since the impedance regulation takes place naturally without requiring amputees’ attention and diminishing their functional capability. Consequently, the proposed interface does not require long training periods or interfere with the control of intact body segments, and is easy to use. Details of the controller and its experimental verification are presented in~\cite{Part2}.

\vspace{.5\baselineskip}

\section{Design of Variable Stiffness Transradial Hand Prosthesis} \label{Sec:DesignofVSAHandProsthesis}

To satisfy the design objectives, an anthropomorphic, VSA integrated, underactuated, compliant hand prosthesis is developed as follows.

\vspace{-.5\baselineskip}

\subsection{Bowden Cable Driven Antagonist VSA} \label{Sec:VSA}

An antagonistic VSA is utilized to control the position and the stiffness of a four-fingered, underactuated, compliant prosthetic hand. Tendon based antagonistic arrangement is preferred since it allows for the elastic elements and actuators to be conveniently placed away from the fingers of the prosthesis. Furthermore, VSA is driven by a Bowden cable based transmission that enables motors, drivers and battery to  be located at any suitable place on the body of amputee. This not only results in a lightweight design, but also enables easy customization of these parts, e.g., force output or battery capacity, for any user.

\subsubsection{Implementation of Antagonistic VSA  using Expanding Contour Cams}

\begin{figure}[htb]
\centering
\resizebox{3.in}{!}{\includegraphics{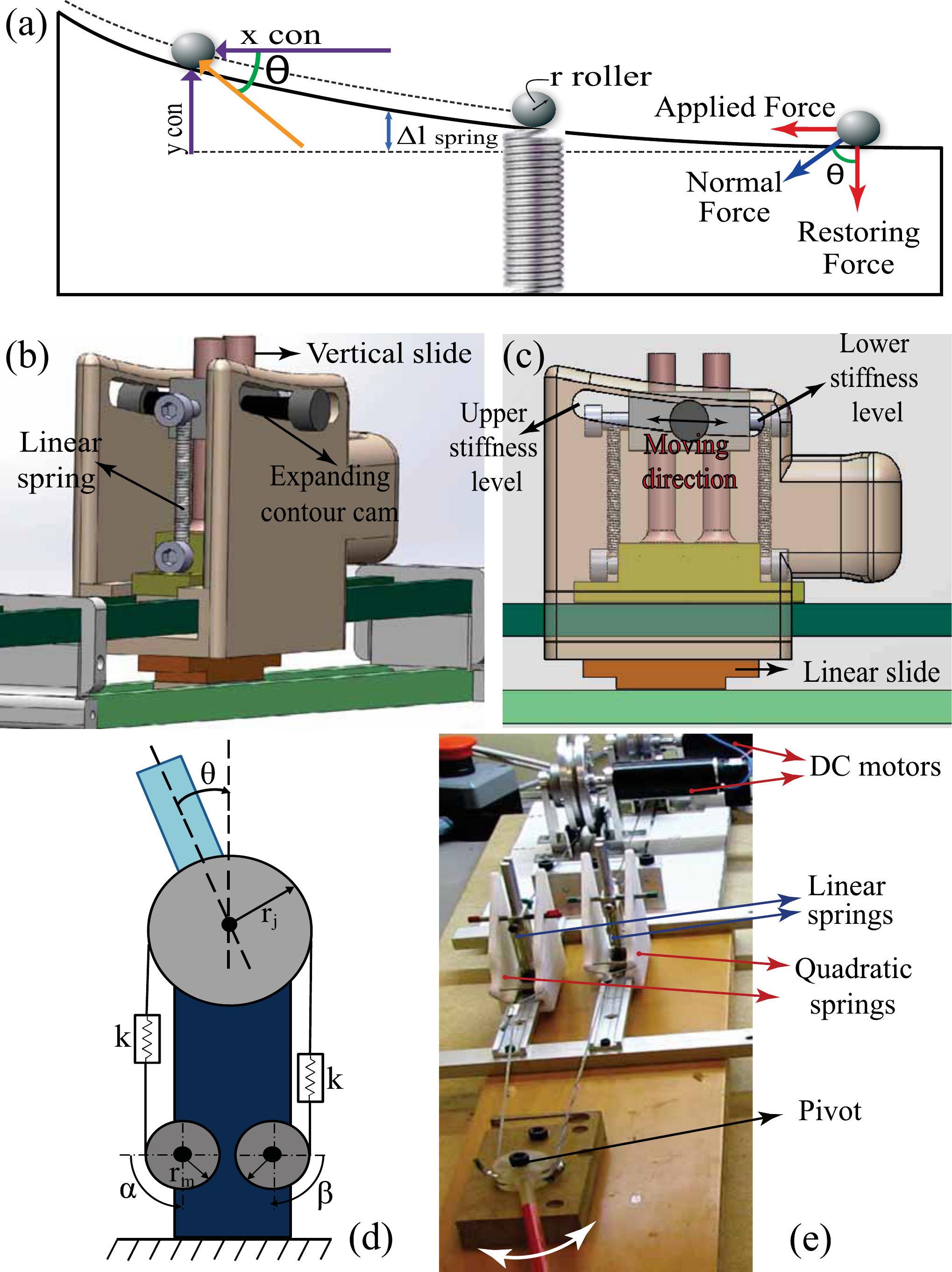}} 
\vspace*{-0.5\baselineskip}
\caption{(a) The expanding contour cam, (b) solid model of the antagonistic VSA, (c) components of VSA, (d) schematic model of the antagonistically driven VSA and (e) the experimental set-up used to verify the VSA.}
\label{quadratic_spring}
\end{figure}

It is well-established that an antagonistic VSA can mimic the independent stiffness and position control of a human limb joint under quasi-static conditions, if the antagonistic spring elements of the VSA have nonlinear (typically quadratic) deflection-force characteristics~\cite{EnglishRussell}. One way of attaining the desired nonlinear spring relationship is to utilize linear springs constrained to move on nonlinear expanding surfaces, called expanding contour cams~\cite{Migliore2007}. In such an arrangement, as shown in Figure~\ref{quadratic_spring}(a-c), when the force is exerted on the system, linear springs extend according to the nonlinear cam surface; hence, a nonlinear relationship between the spring force and the deflection is ensured. The expanding contour cams implement the gradient of the force-deflection relationship and can be designed based on the linear spring constant and the maximum-minimum joint stiffness values.


Since the VSA  aims at modulating the stiffness of prosthetic fingers, the design is implemented based on the maximum  and the minimum  joint stiffness values of human fingers as given in~\cite{Howe1985}. These two design parameters along with the linear spring constant are enough to characterize the shape of the expanding contour as shown below. Different from the design in~\cite{Migliore2007}, our expanding contour cams are designed to be single sided, such that they are compact enough to be integrated into the forearm portion of the prosthesis. Furthermore, the springs on single sided cams are mounted on vertical slides, enabling easy connections, providing stable movements, and preventing the linear springs from bending.

To simplify the design process, the number of parameters required to determine the expanding cam profile are reduced as follows. The free lengths of the springs are selected such that the preload on the linear springs can be set to zero at the thinnest portion of the cam. Since the radius of the rollers is significantly small compared to the cam profile, their effect is neglected in the nonlinear contour equation. Consequently, the force-displacement relationship of elastic elements is chosen to satisfy the quadratic equation
\begin{equation}
\label{fapplied}
F_{app}=ax_{con}^{2}+bx_{con}+c
\end{equation}
\noindent where $F_{app}$ represents the applied force to the roller, $x_{con}$ is the current position of the roller along $x$ axis of the expanding contour, and $a$, $b$ and $c$ are the coefficients of the quadratic equation. Along the $y$ axis, the following relationship holds for the linear springs
\begin{equation}
\label{fspring}
F_{restoring}=k \: y_{con}
\end{equation}
\noindent where $F_{restoring}$ denotes the force applied by the linear spring on the cam along $y$ axis, $y_{con}$ is the current position of the roller along the $y$ axis of the expanding contour and $k$ is the spring constant of linear springs. Neglecting the frictional effects, the cam profile enforces a geometric relation between $F_{app}$ and $F_{restoring}$ that can be expressed as
\begin{multline}
\label{diffeqnsln}
y^{2}_{con}-\bigg(\frac{2a}{3k}\bigg)x^{3}_{con}- \bigg(\frac{b}{k}\bigg)x^{2}_{con} - \bigg(\frac{2c}{k}\bigg)x_{con}-m=0
\end{multline}
Enforcing the following boundary conditions when the linear spring is at the initial point of the expanding contour
\begin{equation}
\label{bouncond}
x_{con}=0, \: y_{con}=0
\end{equation}
\noindent implies that $m=0$. In Eqn.~(\ref{diffeqnsln}), the parameters required to design expanding contour are $a$, $b$, $c$ and the spring constant $k$ of the linear springs. It can be shown that $a$, $b$ and $c$  are directly dependent on the maximum $S_{max}$ and the minimum $S_{min}$ stiffness values of the VSA as follows~\cite{EnglishRussell}
\begin{multline}
\label{stiffness_parameters}
F_{app}=\underbrace{\bigg(\frac{S_{max}-S_{min}}{4r_{j}^{2} \Delta x_{max}} \bigg)}_\text{\textbf{a}} x_{con}^{2} +\underbrace{\bigg(\frac{S_{min}}{2r_{j}^{2}}\bigg)}_\text{\textbf{b}} x_{con} \\
-\underbrace{\bigg(\frac{\Delta x_{max}(S_{max}^{2}-2S_{min}^{2})}{8r_{j}^{2}(S_{max}-S_{min})}\bigg)}_\text{\textbf{c}}
\end{multline}

\noindent where $r_{j}$ denotes radius of the pulley used to implement the VSA and $\Delta x_{max}$ symbolizes the maximum deflection of the linear springs. When the linear springs are unstretched $(x_{con}=0)$, the joint stiffness level is regulated to its minimum level $S_{min}$. In addition to this, when the linear springs reach to their maximum stretch ($x_{con}=x_{max}$), the joint stiffness is regulated at its maximum level $S_{max}$.

\subsubsection{Position and Stiffness Control with Antagonist VSA}

The position and the stiffness of the VSA are controlled through position control of Bowden cables driven by two geared DC motors. Figure~\ref{quadratic_spring}(d) presents a schematic representation of the VSA.  Let $\alpha$ and $\beta$ denote angular position of DC motors, while $S$ and $\theta$ represent joint stiffness and angle, respectively.


Under quasi-static conditions~\cite{Migliore2007,EnglishRussell}, the equilibrium position $\theta$ and  stiffness $S$ of the VSA can be calculated as
\begin{eqnarray}
\theta&=&\frac{r_{m}}{2r_{j}}(\alpha-\beta)-\frac{\tau_{load}}{2r_{j}^{2} (ar_{m} (\alpha+\beta)+b)} \label{thetastiffeqn1}\\
S&=&2ar_{m}r_{j}^{2}(\alpha+\beta)+2br_{j}^{2} \label{thetastiffeqn2}
\end{eqnarray}
\noindent where $r_{m}$ represents the radius of the pulleys attached to the geared DC motors, while the external torque applied to VSA is denoted by $\tau_{load}$.

When control references belonging to the joint position and stiffness are provided, the desired motor positions are computed from Eqns.~(\ref{thetastiffeqn1})-(\ref{thetastiffeqn2}) and the motors are motion controlled to these values. 

\subsubsection{Experimental Verification of VSA}

\begin{figure}[htb]
\centering
\resizebox{3.3in}{!}{\includegraphics{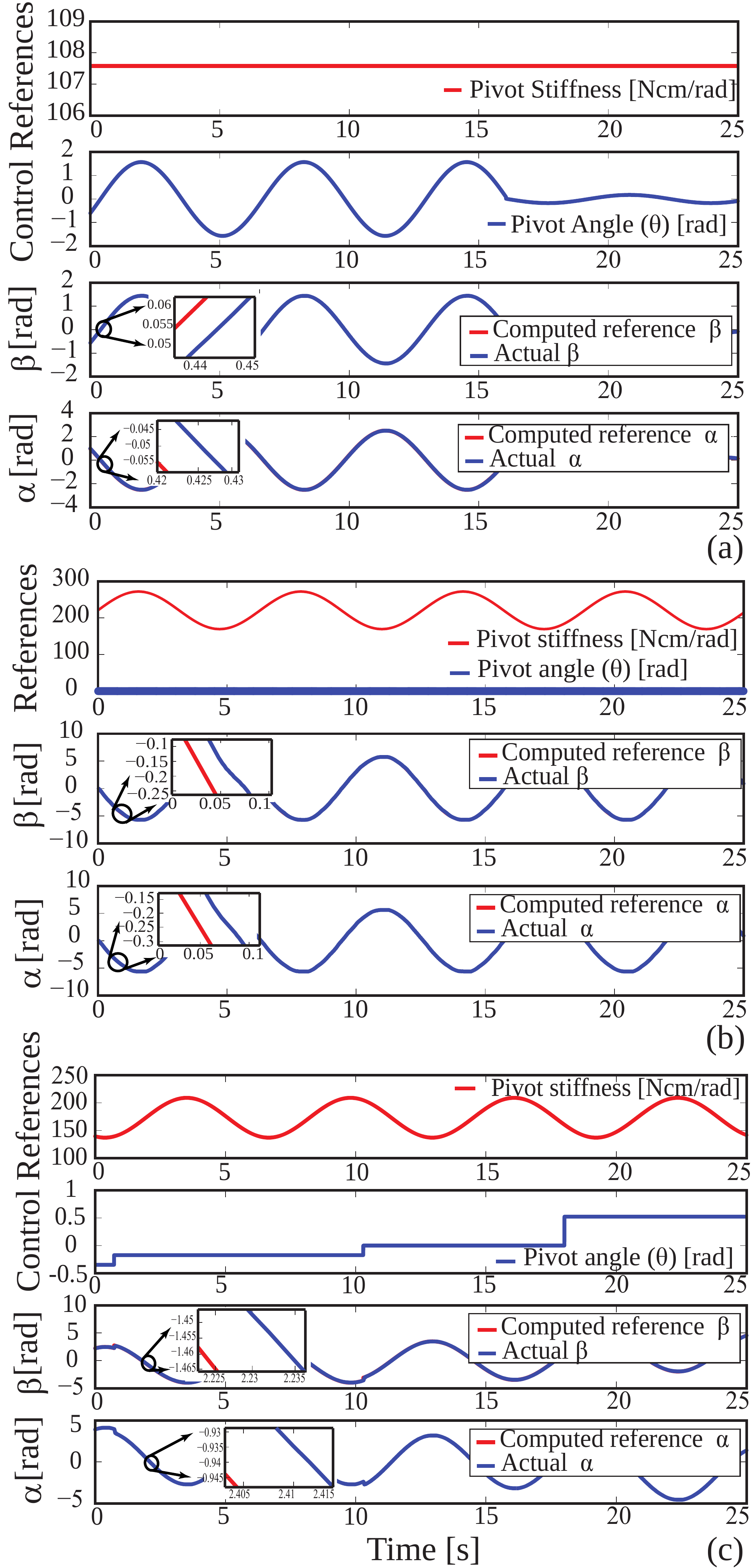}}
\vspace*{-0.5\baselineskip}
\caption{(a) Response to a sinusoidally changing position reference when the stiffness is kept constant. (b) Response to a sinusoidally changing stiffness reference when the position is kept constant. (c) Response to a sinusoidally changing stiffness while the position reference is gradually increased.}
\label{vsa_evaluation_graphs}
\vspace*{-.6\baselineskip}
\end{figure}

To experimentally verify the control performance of VSA, an antagonistic VSA is implemented using expanding contour cams shown in Figure~\ref{quadratic_spring}(b-c) and is connected to a simple pivot. The aim of this experiment is to verify the independent and simultaneous position and stiffness control of the pivot via VSA. Several conditions are tested using the set-up shown in Figure~\ref{quadratic_spring}(e) to evaluate the performance of VSA under realistic conditions. In particular, following three conditions are evaluated sequentially: i) the position of the pivot is kept stationary while its stiffness is changed, ii) the stiffness of the pivot is kept constant  while its position is modulated, and iii) both the position and the stiffness are varied simultaneously. Figures~\ref{vsa_evaluation_graphs}(a-c) present sample results from these experiments.

In Figure~\ref{vsa_evaluation_graphs}(a), the pivot stiffness is kept at a constant level, while the position is simultaneously changed to track a sinusoidal reference of $\pm \pi/2$ rad amplitude for the first 16 seconds and  $\pm \pi/18$ rad amplitude for the rest of the experiment. The results indicate that the position tracking RMS error of the motors are less than $0.003\%$, verifying that the estimated stiffness and position of the pivot can be controlled with $1.2\:10^{-4}\%$ error.

Figure~\ref{vsa_evaluation_graphs}(b) presents results when the pivot stiffness is changed sinusoidally between the intermediate to the high level, while the angular position of VSA is kept constant at zero. The results indicate that the position tracking RMS error of the motors are less than $0.6\%$, verifying that the estimated stiffness and position of the pivot can be controlled with $4.7\:10^{-4}\%$ error.

Figure~\ref{vsa_evaluation_graphs}(c) presents response of the VSA  to a sinusoidally changing stiffness reference, while the position reference is gradually increased with step changes. The results verify that  the position tracking RMS error of the motors are less than $0.004\%$, indicating that the estimated stiffness and position of the pivot can be controlled with $3.8\:10^{-4}\%$ error.

Thanks to robust motion controllers implemented with high gains and at high control rates, the motion tracking error of the motors can be kept low, even under the high friction induced by the Bowden cables. Even though the estimated stiffness and position of the pivot can be controlled with high precision,
there exists other sources of errors, such as unmodelled dynamics of the VSA (control model is valid only under quasi-static conditions) and the elasticity of the cable. However, the experimental results indicate that the position and stiffness tracking performance of VSA are sufficiently high for use in a prosthetic hand, given that the  position and the stiffness references will be provided in a amputee-in-the-loop fashion and at relatively crude discrete levels. In particular,  precise control of these values are not crucial, as just a few discrete (e.g., low, moderate and high) levels of stiffness can significantly improve performance of typical manipulations.


\subsection{Underactuated Power Transmission} \label{Sec:Underactuation}

The proposed hand prosthesis is designed to feature underactuated power transmission, as a means of providing  passive adaptation to various object geometries. Underactuation is preferred as it provides an ideal compromise between dexterous hands that provide versatile and stable grasps at high costs and computational loads, and simple grippers that excel at achieving specific tasks robustly with simple controllers at low costs, but provide relatively small set of grasps. Underactuation is the preferred alternative, since by reducing the number of actuators required to control the system, it not only saves weight, space, and cost, but provides energy efficiency and ease-of-control.

\begin{figure}[htb!]
\centering
\resizebox{3.in}{!}{\includegraphics{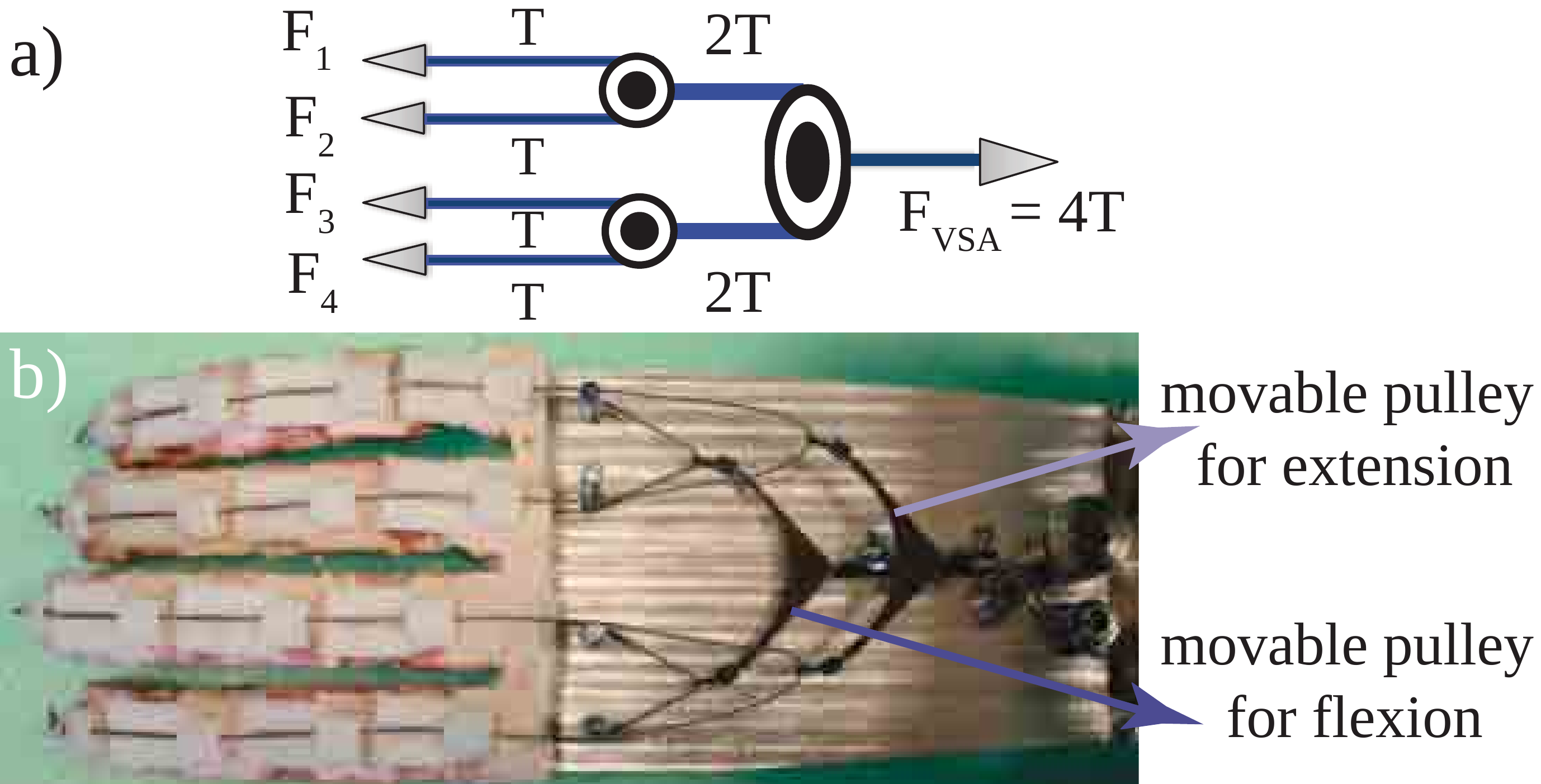}}
\vspace*{-0.5\baselineskip}
\caption{a) Schematic representation of the pulley based transmission that equally distributes tendon tension to each finger, when no external force is acting on the fingers. b) Antagonistic actuation of the hand prosthesis.} 
\vspace{-.25\baselineskip}
\label{underactuationall}
\end{figure}

\smallskip

\subsubsection{Implementation of Underactuated Power Transmission}

The proposed hand prosthesis is designed to be actuated by a single VSA, such that the flexion/extension and the stiffness of fingers are controlled through antagonistic drive tendons. A pulley based power transmission as in  Figure~\ref{underactuationall}(a) distributes tendon tension to each phalanx based on the interaction forces.

When no external force is applied, the tendon tension is transferred equally  to each of the four fingers, where each finger is composed of three compliant joints.  The center of the pivot of VSA can be attached directly to the tendon that flexes the fingers to achieve unidirectional functionality for the hand, where opening of the hand is performed by passive springs. We implement an alternative, where antagonistic springs of VSA are attached to the center of a moveable pulley mechanism, as shown in Figure~\ref{underactuationall}(b). In this arrangement, one of the movable pulleys transmits forces to flex the fingers, while its companion pulley transmits forces to extend them. 

\subsubsection{Experimental Evaluation of the Underactuated Power Transmission}

Natural grasping behaviour of the underactuated prosthesis is tested over different shaped objects. Sample pinch and power grasps are presented in Figure~\ref{illustrative_design}.

Thanks to underactuated power distribution, the compliant fingers naturally adapt to the shape of the objects to ensure that tendon tensions in each finger are equally distributed. That is, each finger seeks appropriate contacts with the object (or a joint limit) to ensure proper force distribution of the tendons. As expected, the grasp type and the motion of each finger depend on the shape of the object and the relative configuration of the prosthesis.

\subsection{Design of Compliant Fingers} 
\label{Sec:compliantfingers}

The proposed hand prosthesis is designed to feature compliant underactuated fingers. Compliant construction results in physical flexibility of the fingers, increasing their adaptability to the  environment and robustness towards impacts. Underactuated kinematics with three compliant joints per finger increases dexterity of the hand, by allowing it to wrap around a wide variety of objects. Underactuation also enables size, weight, and cost
reduction for each finger, since the actuators are typically the largest, heaviest and most expensive components of the device.

The underactuated compliant fingers are designed to mimic the closing sequence of human fingers, such that  a coordinated motion of the phalanges is achieved. In particular, the stiffness of each compliant joint is adjusted such that they maintain the second and third phalanges of the finger in the fully extended configuration until the first phalanx comes in contact with an obstacle or reaches its mechanical limit. When the mechanism is free of contacts and within joint limits, it behaves like a single rigid body. But when the motion of a phalanx is resisted, the force generated by the tendon overcomes the spring preload and the adjacent phalanx initiates motion. The motion continues sequentially until movements of all phalanges are resisted, due to either contacting with an object or reaching to the joint limits. Hence, each compliant finger is capable of producing many of the natural finger trajectories of a human hand and the tendon force is properly distributed over all phalanges.
\smallskip

\subsubsection{Material Selection}

Selection of appropriate material to cover the rigid phalanges and compliant joints is important to achieve robust fingers with a soft delicate touch. Selection of a high viscous silicon rubber results in high force requirements from the tendon and increases the energy consumption during bending, while selection of a less viscous silicon rubber causes easy cracking of the material, decreasing the robustness of the finger. After testing many different materials, SILASTOSIL\textregistered~28-700 FG is evaluated to provide the best compromise among polyurethane materials to serve as the base material for the underactuated compliant fingers. 

\begin{figure*}[t]
\centering
\resizebox{5.5in}{!}{\includegraphics{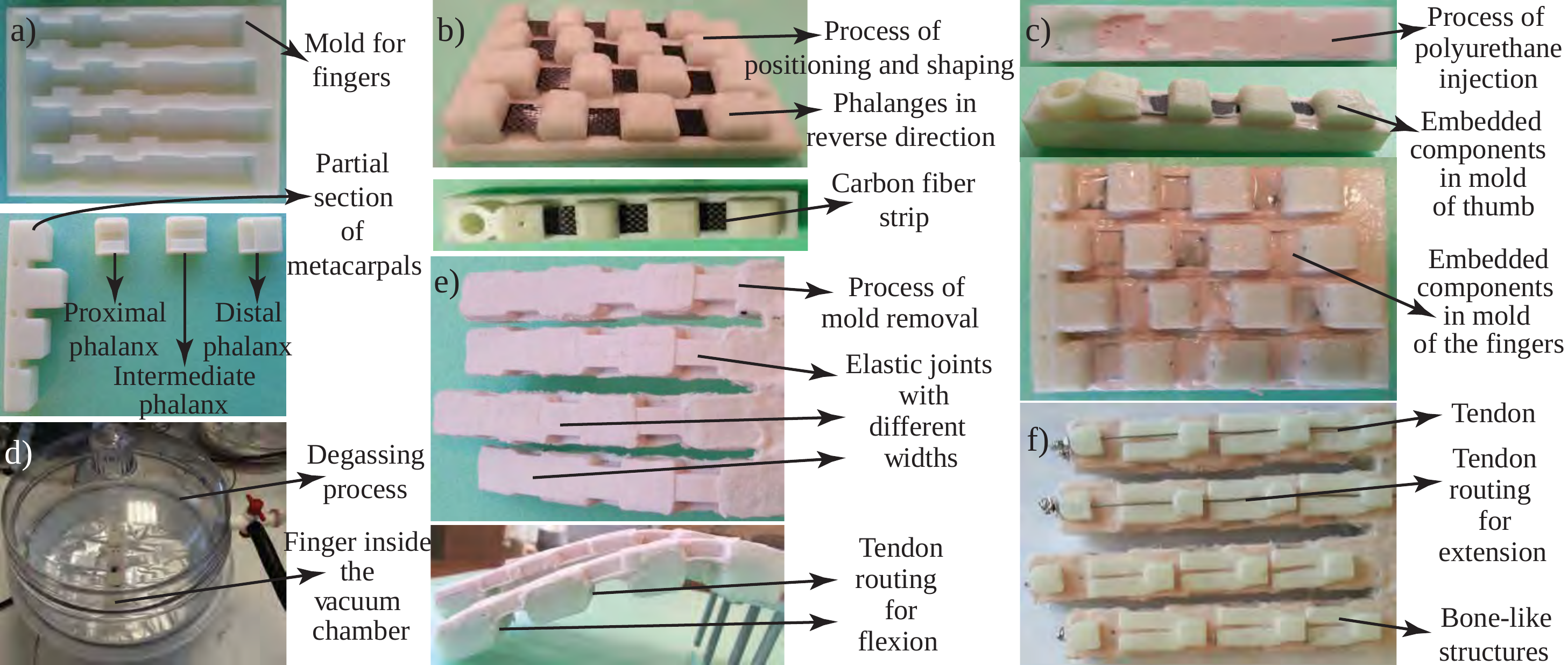}}
\vspace*{-0.35\baselineskip}
\caption{The six step process used to fabricate the compliant fingers.}
\label{fabrication}
\vspace*{-0.85\baselineskip}
\end{figure*}

Implementation of proper stiffness for the compliant joints play a crucial role in achieving the desired coordinated motion of the anthropomorphic fingers. Phalanges are fabricated with ABS plastic through rapid prototyping, while the compliant joints are fabricated using the polyurethane material. While polyurethane material is proper to serve as the base material to cover the compliant fingers, including the phalanges and compliant joints, it is necessary to further adjust the stiffness of each compliant joint to implement the desired coordinated motion and to anisotropically strengthen these joints against twisting and bending. Along these lines, carbon fiber strips are embedded inside each compliant joint. Lightweight carbon fiber strips not only act as leaf springs used to implement desired level of compliance at each joint, but also help support the joints by structurally reinforcing them against twisting and undesired bending forces.

\vspace{2mm}
\subsubsection{Fabrication of the Compliant Fingers}

Fabrication of compliant fingers consists of several stages as presented in Figure~\ref{fabrication}. In the first step,  rigid parts, such as parts of phalanges and molds, are fabricated using additive manufacturing as shown in Figure~\ref{fabrication}(a). Additive manufacturing  enables intricate design of phalanges and finger molds to be custom built for each amputee. Furthermore, accuracies of 100 $\mu$m are easily achievable at low manufacturing costs.

In the second step, finger molds are used for precise arrangement of phalanges, compliant joints and fingers. Epoxy resin infused carbon fiber sheets are  incorporated into the design as in Figure~\ref{fabrication}(b), before injecting silicone rubber in liquid form. This way, even though each joint consists of same materials, silicone rubber molds with varying widths enable specific stiffness levels to be associated at each joint (see Figures~\ref{finger_specs_crosssection_strain}(c-d)). In particular, the stiffness is increased from proximal to distal joints. Consequently, the joint flexion initializes at the metacarpophalangeal joint, continues at the proximal interphalangeal joint, and finalizes at the distal interphalangeal joint.

In the third step, highly-adhesive silicone rubber is injected to the mold in its liquid form as depicted in Figure~\ref{fabrication}(c). Before pouring the silicon resin, routing holes at phalanges are strapped in order to prevent polyurethane flow inside tendon routes. In addition, a release agent is used to avoid bonding of cured polyurethane to mold surfaces and to facilitate the releasing of the part from the crinkled-shaped mold.

As a consequence of injecting silicon rubber into the mold, air bubbles  occur inevitably, which may cause inhomogeneous material distribution, adversely affecting the stiffness of each joint; hence, the coordinated motion of the finger. In the fourth step, degassing is implemented with -0.2 to -0.5~bar pressure to remove air bubbles from the silicone material, as presented in Figure~\ref{fabrication}(d). The recommended cure time for the resin is about 12~hours at the room temperature and the complete molding process takes about 13~hours.

In the fifth step, the fingers are removed from the molds and tendons responsible for force transmission are inserted into their routes, as shown in Figure~\ref{fabrication}(e).

In the final step shown in Figure~\ref{fabrication}(f), bone-like structures, as in Figure~\ref{finger_specs_crosssection_strain}(b), are placed on the upper surface of fingers to induce physical joint limits. These structures constrain the finger extension after it reaches its fully extended horizontal position, preventing fingers to bend in the reverse direction. High friction soft finger pads produced using the silicone material are added on the contact surfaces of phalanges to improve slip resistance of the fingers~\cite{softmaterials,skinmaterials}, as in Figure~\ref{finger_specs_crosssection_strain}(a).

\subsubsection{Cable Routing for the Compliant Fingers}

\begin{figure}[htb!]
\centering
\resizebox{3.35
in}{!}{\includegraphics{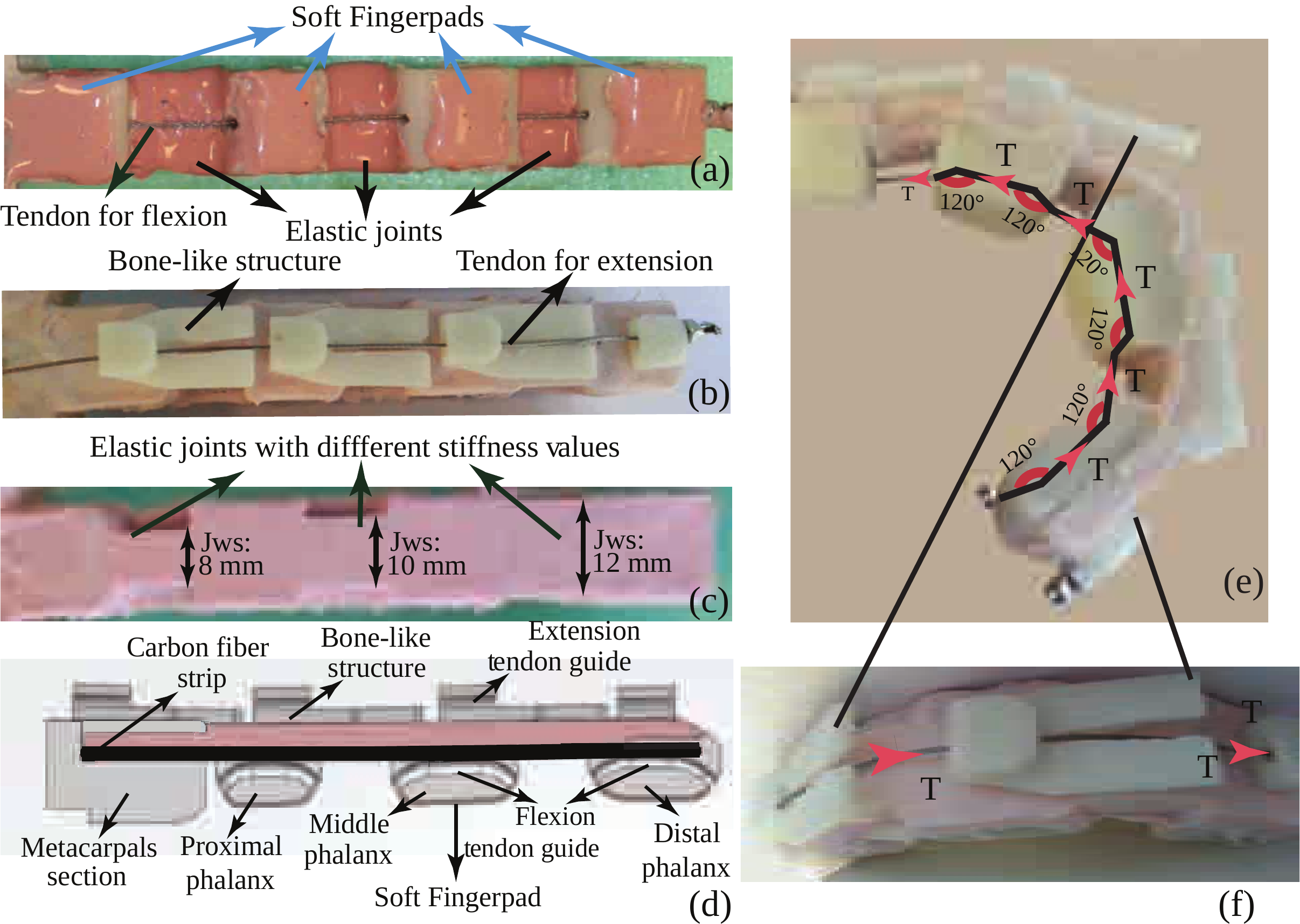}}
\vspace*{-0.35\baselineskip}
\caption{(a) Bottom view with soft finger pads. (b) Top view with bone-like joint limits. (c) Side view with varying joint widths resulting in customized joint stiffness. (d) Cross-section of a solid model of the compliant finger with cable routing. (e) Flexion tendon routing. (f) Extension tendon routing.}
\vspace*{-.5\baselineskip}
\label{finger_specs_crosssection_strain}
\end{figure}

Force transmission of the compliant fingers is achieved through the flexion and the extension tendons. Due to the inherent stiffness of each compliant joint, higher forces are required while closing the fingers. To facilitate easier closing of the fingers, flexion cable channels are implemented with 120$^{\circ}$ angles, such that larger moment arms are implemented for the flexion, increasing the moments acting on the phalanges. Figure~\ref{finger_specs_crosssection_strain}(d) depicts the cross-section of solid model of a compliant finger, while Figures~\ref{finger_specs_crosssection_strain}(e-f) show the flexion and extension tendon routing on a finger prototype.

\subsection{Implementation of the Transradial Hand Prosthesis} 
\label{Sec:Assembly}

\begin{figure}[htb]
\centering
\resizebox{2.75in}{!}{\includegraphics{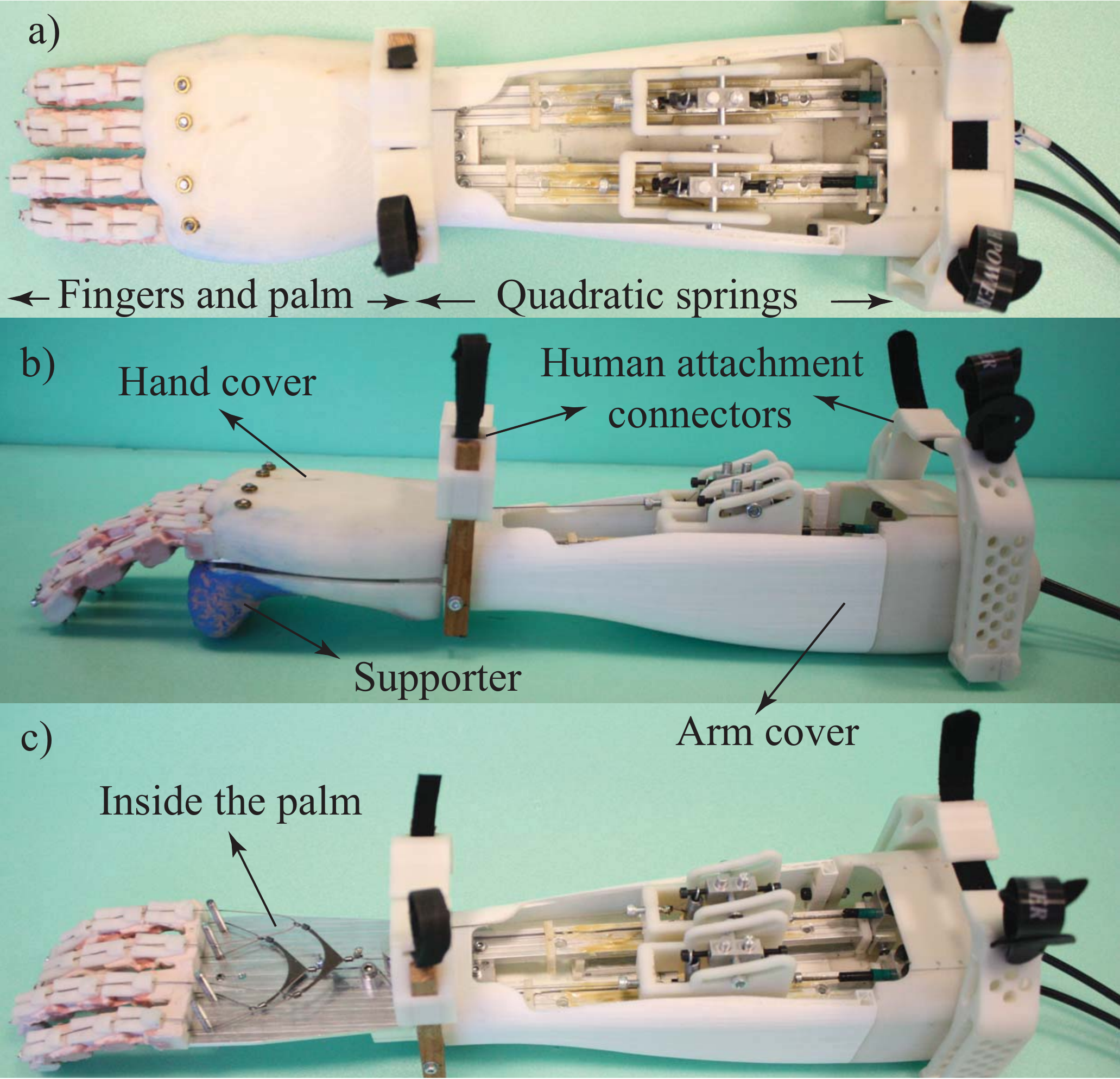}}
\vspace*{-.35\baselineskip}
\caption{(a) Top view,  (b) side view, and (c) interior of the transradial hand prosthesis prototype}
\vspace*{-.75\baselineskip}
\label{hand3d}
\end{figure}

The hand prosthesis consists of three main components: VSA, forearm, and compliant fingers. The expanding contour cams of VSA, the forearm in which the nonlinear springs of VSA are embedded on linear sliders, and device covers are fabricated through additive manufacturing, while the palm on which compliant fingers are attached is constructed using a laser cut aluminum sheet. Figure~\ref{hand3d} presents an assembled prototype. 

Since the actuators, motor drivers and batteries of the Bowden cable driven VSA can be placed remotely anywhere on the body, e.g., can be kept inside a backpack, they are not integrated inside the prosthesis. This decision helps keep the device weight and cost low, as it provides extensive design flexibility while choosing/customizing these components according to the needs of the amputee. Note that, integration of these component into the forearm is possible, but induces challenging size and weight constraints for these components.


The specifications required from the drive train are characterized as follows. The maximum force required on the drive tendon such that a 1.5~kg object can be lifted at the maximum stiffness level is verified as 160~N. The force required to close the hand at the lowest stiffness level with no object in the hand is characterized as 20~N. The maximum speed required on the drive tendon is characterized as 20~mm/s, such that all fingers reach their joint limits within 2 seconds, when the fingers are free to move.

Geared DC motors that satisfy the maximum force and speed requirements are selected to implement the drive train. With these motors, the minimum and the maximum stiffness of the drive tendon are experimentally characterized as 135~Nmm/rad and 545~Nmm/rad, respectively. The overall weight of the device (excluding the motors, drivers and battery) is 1.1~kg, which is lower than the natural weight of the corresponding part of human limb~\cite{armweight1,armweight2}. Light weight may cause less fatigue for the amputee, and if necessary, the weight of the prosthesis can easily be adjusted to match the weight of the lost limb. Note that the current research prototype is developed for verifying functionality and has not been optimized for size and weight. The drive train specifications of the hand prosthesis are presented in Table~\ref{handspecs}.

\begin{table}[htb]
\centering \small
\caption{Technical Specifications}
\label{handspecs}
 \vspace*{-0.5\baselineskip}
\begin{tabular}{l|l}
\hline \hline
      {Maximum Tendon Force} & 160 N   \\ \hline
      {Minimum Tendon Force} & 20 N   \\ \hline
      {Maximum Tendon Speed} & 20 mm/s \\ \hline
      {Minimum Joint Stiffness} & 135 Nmm/rad    \\ \hline
      {Maximum Joint Stiffness} & 545 Nmm/rad  \\ \hline
      {Weight} & 1.1 kg    \\ \hline
\end{tabular}
\vspace*{-1\baselineskip}
\end{table} \normalsize

To select a proper sized battery pack, the power consumption of the transradial hand prosthesis is experimentally characterized, as it is dependent on the grasp type and the friction losses in the system. Power and pinch grasps are studied as these are most commonly used. Each grasp type is repeatedly executed on a wide variety of objects, when the hand stiffness is set to high and low values, respectively. Furthermore, the average power consumed while modulating the hand stiffness from its lowest level to highest level is also characterized. The mean power consumption for each grasp type and stiffness change are presented in Table~\ref{energy}.

\begin{table}[htb]
\centering \small
\caption{Power Consumption}
\label{energy}
 \vspace*{-0.5\baselineskip}
\begin{tabular}{ c|c|c }
\hline \hline
\multicolumn{3}{c}{Energy Requirements [mWh]} \\
\hline
     & Low Stiffness & High Stiffness \\ \hline
      {Power Grasp} & 30  &  81  \\ \hline
      {Pinch Grasp} & 18  &  75  \\ \hline
      Stiffness Modulation & \multicolumn{2}{c}{4.9}\\  \hline
\end{tabular} \normalsize
\vspace*{-0.15\baselineskip}
\end{table}

The highest energy consumption takes place during a power grasp when the stiffness is simultaneously modulated from low to high. The results indicate that a 50~g rechargeable LiFePo4 battery pack with 1500~mAh enables execution of this grasp for 120 times. Furthermore, these battery modules can be charged within an hour. Number of battery modules integrated into the system can be personalized based on the needs and preferences of the amputee.

\section{Experimental Evaluation of the Variable Stiffness Transradial Hand Prosthesis}
\label{Sec:Experiments}

The aim of the experimental evaluation is to reveal the feasibility of the working principle of the proposed hand prosthesis. A set of experiments is conducted in order to validate the independent position and stiffness modulation of the hand prosthesis. The experiment consists of two tasks, where the objective of the first task is to verify the controllability of the stiffness when the position is regulated at a constant value, while the second task aims to verify that the desired position can be modulated when the stiffness parameters are kept constant.

\subsection{Experimental Setup and Procedure}

\begin{figure}[htb!]
\centering
\resizebox{3in}{!}{\includegraphics{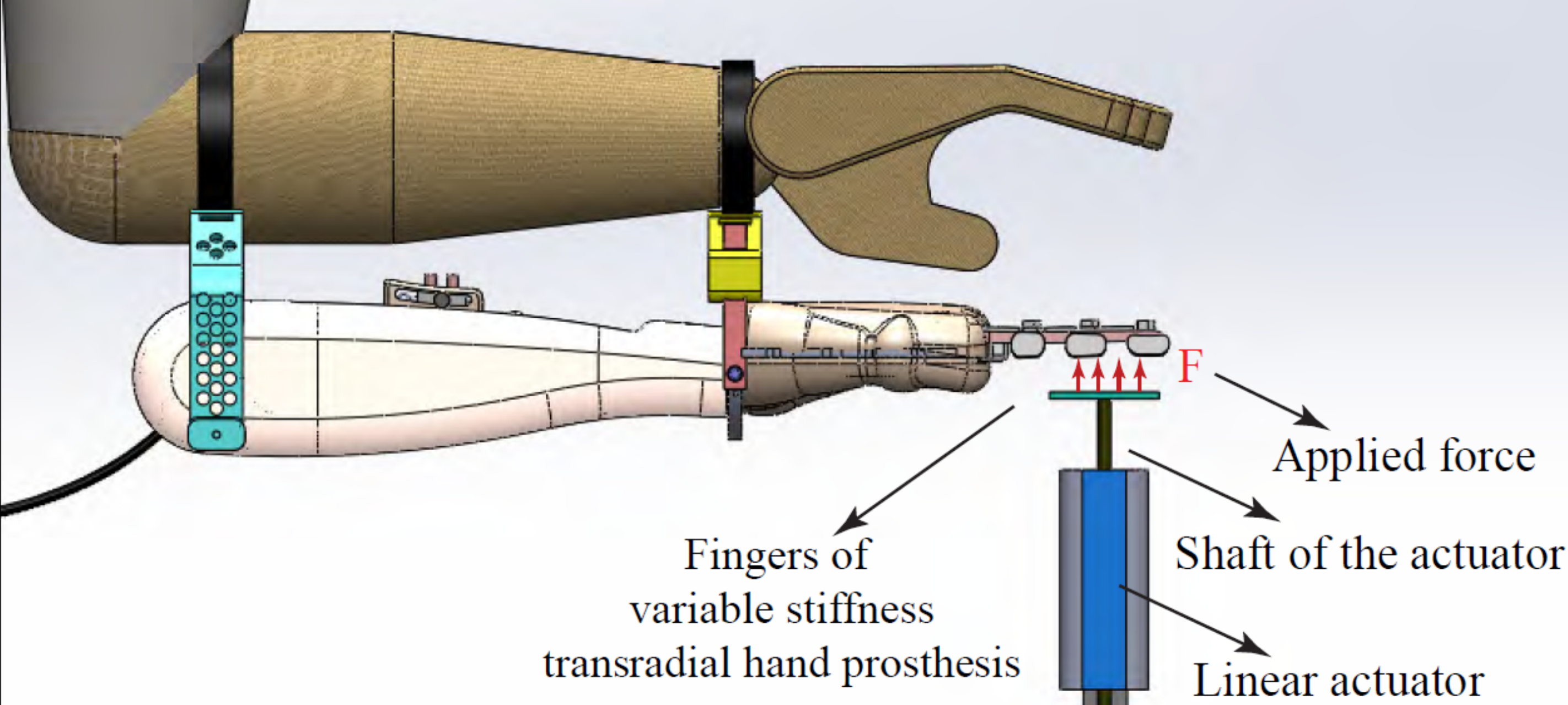}}
\vspace*{-.5\baselineskip}
\caption{Schematic representation of the experimental setup.}
\vspace*{-.25\baselineskip}
\label{exp1_shematic_part1}
\end{figure}

The experimental setup consists of a direct drive linear actuator with a built-in high resolution incremental encoder, placed under the four fingers of the transradial prosthesis, as shown in Figure~\ref{exp1_shematic_part1}. During the experiments, the gravitational force acting on the actuator is compensated with a counter mass, while the actuator is force controlled. 
All controllers are implemented in real-time at 500~Hz with a PC workstation equipped with a DAQ card. In these experiments, the position and the stiffness levels are not controlled by the volunteers, as the goal is to perform a verification of the hand prosthesis independent of its user interface. Hence, during these experiments, the reference values for the position and the stiffness are set by the PC workstation.

The experiment is composed of two tasks with 10 repetitions for each condition of each task. During the first task, the position of the VSA is kept constant at $0^{\circ}$, that is, the angular position of the metacarpophalangeal (MCP) joint  is set to $0^{\circ}$, while the stiffness of VSA is adjusted to three distinct stiffness values that correspond to low, intermediate, and high stiffness levels for the fingers. The stiffness of the fingers are experimentally determined by applying a linearly increasing force to flex the fingers and recording their deflection.

During the second task, the stiffness of the VSA is kept constant at its intermediate level, while the position of the VSA is adjusted to three distinct position values that correspond to low, intermediate, and high flexion of the fingers. The position of the fingers is determined by recording the position of the linear actuator under zero force control, while the stiffness of the fingers are determined by applying a constant force to resist flexion the fingers at the equilibrium position and recording the resulting deflection.

\subsection{Experimental Results}

\begin{figure}[htb]
\centering
\resizebox{3.2in}{!}{\includegraphics{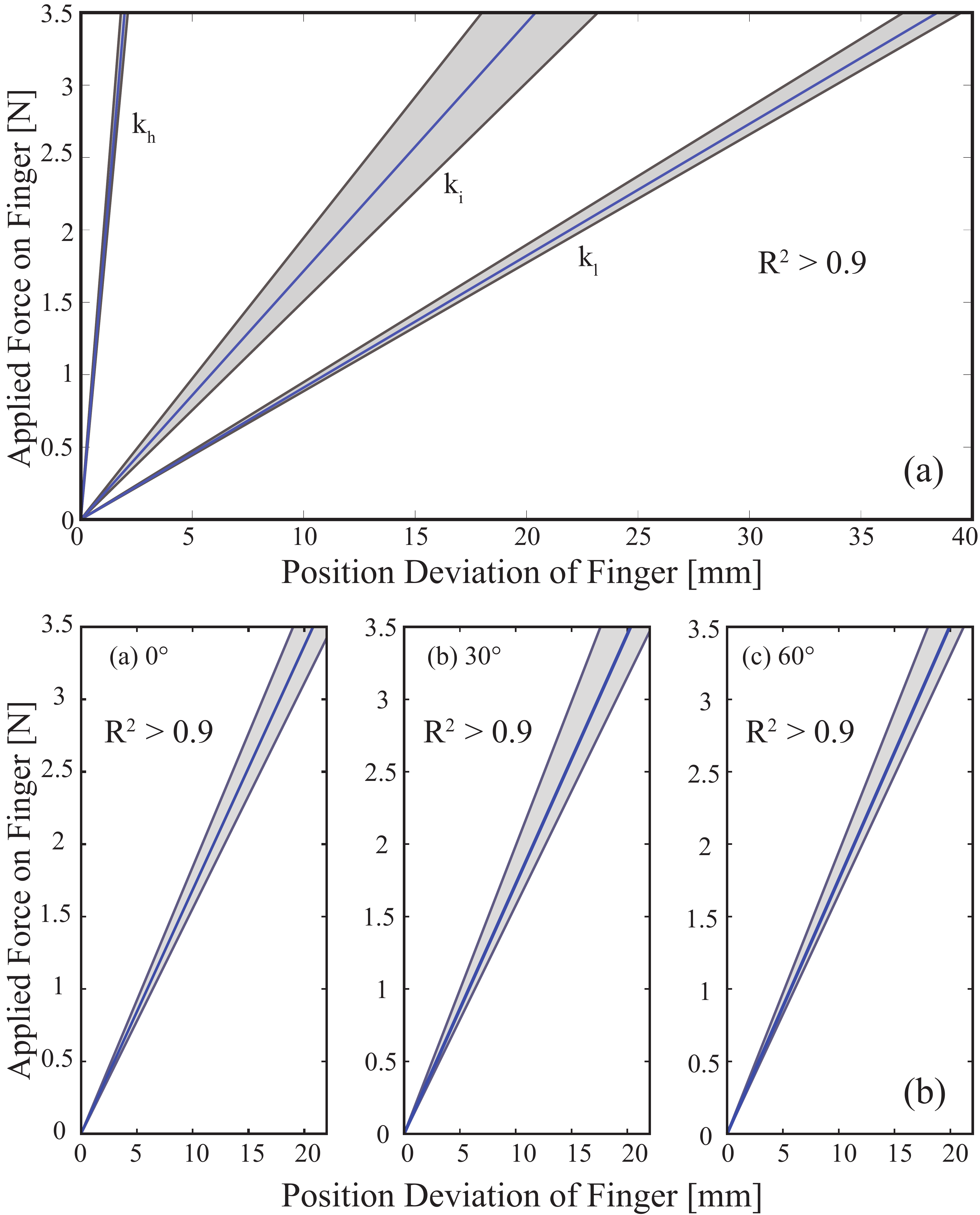}}
\vspace*{-.25\baselineskip}
\caption{(a) Stiffness modulation of hand prosthesis. Gray zone presents the best linear fit of each trial and the blue line presents the average value of ten trials. (b) Position control of hand prosthesis. Gray zone presents the results of each trial. The blue line presents the average value of ten trials.}
\label{st_po_wo_emg}
\vspace*{-0.25\baselineskip}
\end{figure}

Figure~\ref{st_po_wo_emg}(a) presents the experimental results for the case when VSA is adjusted to three distinct stiffness values that correspond to low, intermediate and high stiffness levels for the fingers, while the finger positions are kept constant. In particular, shaded regions represent all the linear fits recorded for 10 trials, while the dark line represent their mean. The slopes of these lines indicate that low, intermediate and high stiffness for the fingers are
$k_{l}$=0.091~N/mm,  $k_{i}$=0.17~N/mm, and $k_{h}$=1.8~N/mm, respectively. The $R$ values for these linear fits are evaluated to be higher than 0.98.


Figure~\ref{st_po_wo_emg}(b) presents the experimental results for the case when VSA is adjusted to intermediate stiffness level, while the finger positions, that is, the angle between the MCP joint of the fingers and the palm surface, are regulated to $0^{\circ}$, $30^{\circ}$, and $60^{\circ}$, respectively. Once again, the shaded regions represent all the linear fits recorded for 10 trials, while the dark line represent their mean. The slopes of these lines indicate that stiffness level of the fingers are $k_{0^\circ}$=0.17~N/mm,  $k_{30^\circ}$=0.17~N/mm, and $k_{60^\circ}$=0.18~N/mm, respectively. The $R$ values for these linear fits are evaluated to be higher than 0.98.


Experimental results provide strong evidence that the stiffness and the position of the transradial hand prosthesis can be controlled independently, with high repeatability while executing predefined tasks. The impedance characteristics of the compliant fingers of the VSA prosthesis closely match the characteristics of human fingers as presented in~\cite{Howe1985}. The  characterization results are also compatible with the results presented in~\cite{finger_extension}, as the flexion/extension movements performed by an anatomically human-like robotic index finger necessitates similar amount of muscle forces.

\begin{figure*}[htb]
\centering
\resizebox{7.1in}{!}{\includegraphics{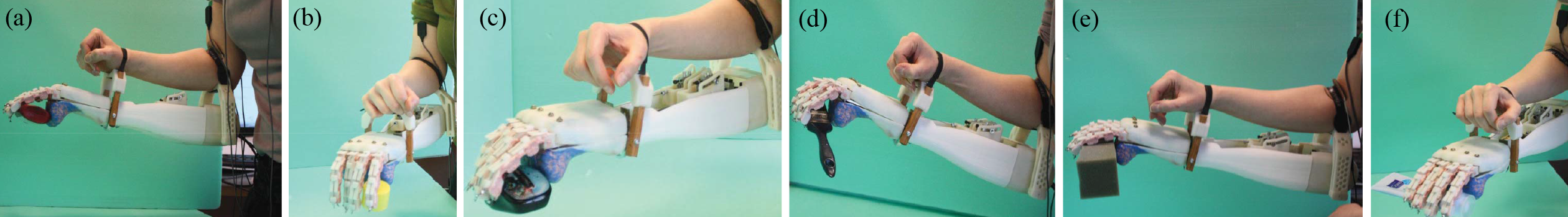}}
\vspace*{-0.75\baselineskip}
\caption{Demonstration of variable stiffness transradial hand prosthesis performing various grasps while interacting with (a) a glue box, (b) a bottle cap, (c) a mouse, (d) a brush, (f) a sponge, and (g) a cream tube.}
\vspace*{-0.5\baselineskip}
\label{illustrative_design}
\end{figure*}

\section{Illustrative Experiments and Discussions}
\label{Sec:IllustrativeExperiments}

Given that only the position and the stiffness of the driven tendon are directly regulated by the VSA, in general, the resulting position and stiffness of the fingers depend on the interaction. To test the usefulness of the variable stiffness transradial hand prosthesis, the device is attached to six volunteers, as shown in Figure~\ref{exp1_shematic_part1}. All volunteers signed informed consent forms approved by the IRB of Sabanci University. The volunteers were given the control of the position and the stiffness of the prosthesis through the sEMG based tele-impedance controller~\cite{Part2,hocaoglue12}.

In particular, sEMG signals measured from the surface of the upper arm, chest and shoulder were used to automatically adjust the stiffness level of the prosthesis to that of the upper arm, while the position regulation was intentionally controlled by the volunteers by moving their shoulder muscles. With this natural control interface, the volunteers were asked to grasp 16 objects with a wide variety of shapes (e.g., rectangular, elliptic, complex) and  compliance levels (e.g., rigid, elastic). Participants were asked to grasp and hold the objects for a while, then release them back onto the surface. In the same manner as in Section~\ref{Sec:Experiments},  the required stiffness level of VSA to safely grasp each object depended on the interaction. The volunteers were successful at modulating their impedance and grasping a wide variety of objects with the sEMG interface, as shown in Figure~\ref{illustrative_design}. Videos demonstrating several illustrative grasps are available at \url{https://youtu.be/fGFIKSSmtDg}.

The current prototype emphasizes simplicity, ease of use and adaptability; hence, implements a two-degree-of-freedom underactuated power transmission to allow for the position and stiffness change of the prosthetic hand. Successful employment of the prosthesis depends on the amputee making proper decisions on how to interact with the object. Our extensive experiments with volunteers indicate that humans are very skillful at learning how to interact with the environment with such a device. On average, it took $3.2 \pm 1.3$~minutes for a volunteer to get used to the device and successfully complete the required manipulation tasks.

The time elapsed for grasping and releasing of the 15 objects in the video are calculated as $1.218\pm0.564$~sec and $0.819\pm0.48$~sec, respectively. The time required to make a fist is about 2 sec. The grasping performance of the proposed prosthesis prototype is comparable to commercial ones~\cite{openingclosingtime}.

The robustness of the hand prosthesis is also tested during the user studies. During the user studies, volunteers repeatedly impacted the fingers to various surfaces. Compliant fingers made of silicon rubber and ABS material were robust to such impacts and did not sustain any damage.

In the current research prototype, a passive support is preferred to oppose the fingers, instead of an active thumb. This decision helps keep the system and the controller simple. Our experiences with the volunteers indicate that the passive support is adequate for implementing a wide variety of functional grasps. During post trial interviews, none of our volunteers complained about the functionality of this support.

\section{Conclusion and Future Work}

In this study, the design, implementation and experimental evaluation of a low-cost, customizable, easy-to-use variable stiffness transradial hand prosthesis have been presented. User studies indicate that the device is dexterous enough to successfully interact with a wide variety of environments.

The main goal of any new prothesis is to provide a fulfilling functionality to amputees, decreasing or even reversing the relatively high abandonment rate for current high-tech prosthetics. While some of the commercial devices provide many extra functions including a wrist rotation, these devices are complicated and require special training and long rehabilitation periods, which discourages a great majority of amputees to continually use such devices.

Achieving a dexterous, anthropomorphic, adaptable, robust, low-cost mechatronic system design is only the half of the story towards achieving an ideal prosthesis, where design of a natural and easy-to-use control interface for the mechatronic system is the other half. We present such a human-machine interface, called \emph{tele-impedance control}, in~\cite{Part2}.


It is possible to add a thumb to the tendon driven power transmission; however, substantial effort is required for the intricate control of thumb orientation and stiffness,such that it becomes functional and enhances the grasping performance. Alternatively, to achieve an anthropomorphic look with an aesthetically pleasing appearance, a compliant unactuated thumb can be also added to the system. Furthermore, prosthesis gloves can be worn over the transradial hand prosthesis to achieve the natural appearance of human skin.

Our future work will  focus on improving the anthropomorphism by inclusion of a functional thumb to the hand prosthesis, conducting more extensive testing with amputees, and refinement of the prosthesis  based on the feedback collected from amputees.


\section*{Acknowledgment}
This work has been partially supported by Tubitak Grants  111M186, 115M698 and Marie Curie IRG Rehab-DUET. 

\ifCLASSOPTIONcaptionsoff
  \newpage
\fi



%

\bibliographystyle{IEEEtran}
\bibliography{mymanuscript_d}

\begin{IEEEbiography}[{\includegraphics[width=1in,height=1.25in,clip,keepaspectratio]{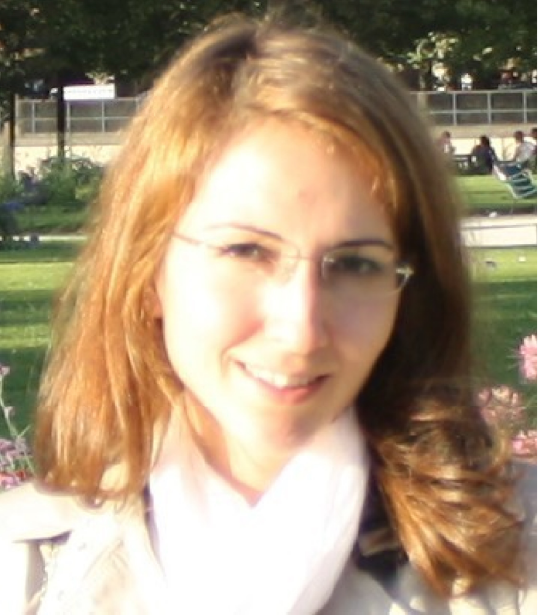}}]
{Elif Hocaoglu} received her Ph.D. degree in Mechatronics Engineering from the Sabanci University, Istanbul in
2014. She worked as a post doctoral researcher at Advanced Robotics Laboratory at Italian Institute of Technology and then Human Robotics Laboratory at Imperial College London. Currently, she is an assistant professor at Istanbul Medipol University. Her research is in the area of physical human-machine interaction, in particular sEMG based control interfaces with applications to upper extremity prostheses, rehabilitation and assistive devices. Her research extends to wearable robotics.
\end{IEEEbiography}

\begin{IEEEbiography}[{\includegraphics[width=1in,height=1.25in,clip,keepaspectratio]{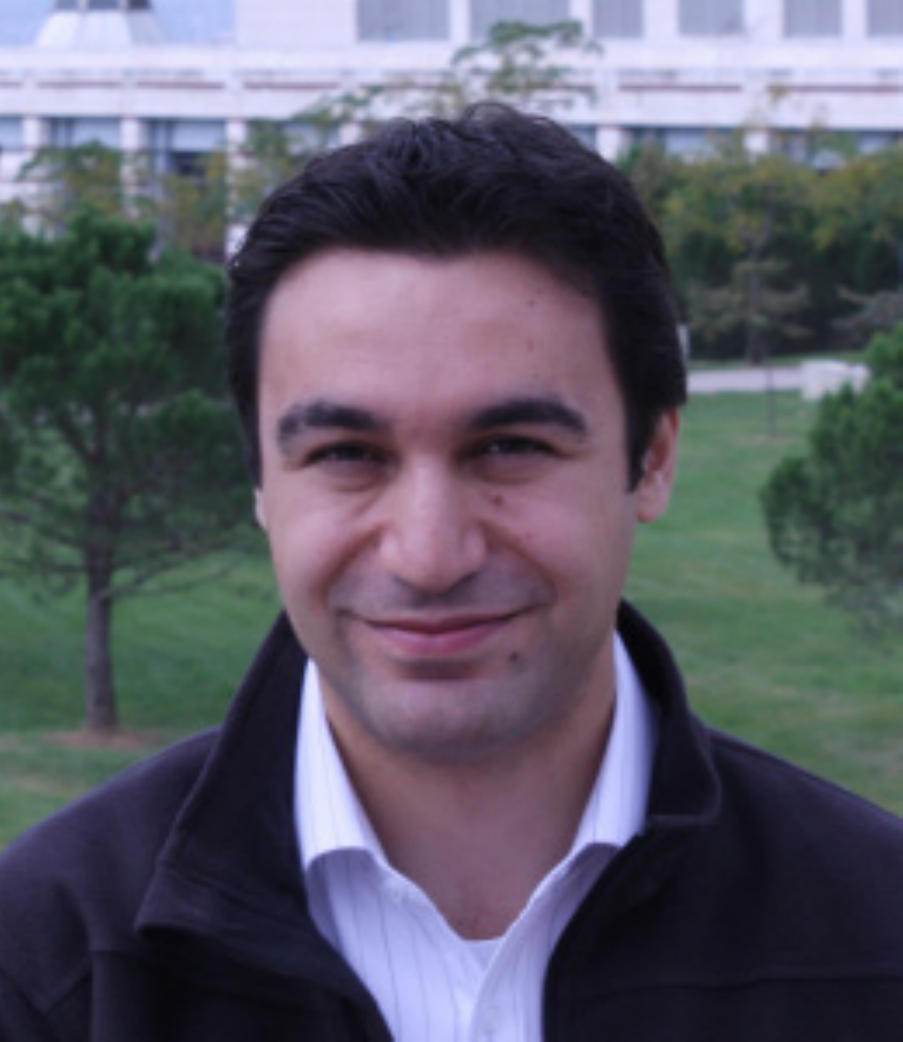}}]
{Volkan Patoglu} received his Ph.D. degree in
Mechanical Engineering from the University of Michigan, Ann Arbor in
2005. He worked as a post doctoral research associate in
Mechatronics and Haptic Interfaces Laboratory at Rice University.
Currently, he is a professor at Sabanci University. His
research is in the area of physical human-machine interaction, in
particular, design and control of force feedback robotic systems
with applications to rehabilitation and skill training. His research
extends to cognitive robotics. 

Dr. Patoglu has served as an associate editor for the IEEE Transactions on Haptics (2013--2017)
and is currently an associate editor for the IEEE Transactions on Neural Systems and Rehabilitation Engineering and IEEE Robotics and Automation Letters.
\end{IEEEbiography}

\vfill


%
%

\end{document}